\documentclass[preprintnumbers,twocolumn,showpacs,amsmath,amssymb,floatfix,9pt,prd,
superscriptaddress,nofootinbib]{revtex4}
\usepackage{latexsym}
\usepackage{epsfig}
\usepackage[ansinew]{inputenc}
\usepackage{amsmath}
\usepackage{amssymb}
\usepackage{graphicx}
\usepackage{color}
\usepackage[colorlinks]{hyperref}

\begin{document}

\title{ A new relativistic model of hybrid star with interactive quark matter and dense baryonic matter}

\author{Koushik Chakraborty}
\email{koushik@iucaa.ernet.in } \affiliation{Department of Physics,
Government Training College, Hooghly - 712103, West Bengal, India}

\author{ Farook Rahaman}
\email{  rahaman@iucaa.ernet.in} \affiliation{Department of
Mathematics, Jadavpur University, Kolkata 700 032, West Bengal,
India}
\author{Arkopriya Mallick}
\email{arkopriyamallick@gmail.com} \affiliation{Department of Mathematics,Jadavpur University, Kolkata - 700 032 West Bengal, India}

\date{today}

\begin{abstract}
We propose a relativistic model of hybrid star admitting conformal symmetry considering quark matter and baryonic matter as two different fluids. We define interaction equations between the normal baryonic matter and the quark matter and study the physical situations for repulsive, attractive and zero interaction between the constituent matters.  From the interaction equations we find out the value of the equation of state (EOS) parameter for normal baryonic matter which is found to be consistent with the value obtained from the Walecka model for nucleonic matter at high density. The measured value of the Bag constant is used to explore the space time geometry inside the star. The theoretical mass-radius values are compared with the available observational data of the compact objects. From the nature of the match with the observational data, we predict the nature of interaction that must be present inside the hybrid stars.

\end{abstract}


\maketitle

\section{Introduction}
~~~~The general conjecture is that gravitational collapse of massive stars ( $M > 8 M_{\odot}$ ) after a Type II supernova leads to neutron stars. \cite{bethe} These dense compact stars have been an object of various physical speculations due to possible extreme conditions of density and pressure at the core of the star. In fact, the physical nature of the interior of neutron star is still a point of debate among the researchers. In spite of the availabilty of plenty of observational data, the problem is hardly resolved.
The extreme conditions of pressure and density inside the neutron star may lead to the phase transition of neutrons to hyperons, bosons, free quark matter or strange quark matter. Cameron \cite{cameron}predicted that hyperons must be produced inside the neutron star. At high densities some of the nucleons may be converted to hyperons due to weak interaction as that is energetically more favourable. However, quark matter may also be present in the interior of the neutron star. Due to extreme asymptotic momentum transfer and density at the core of the neutron star, quarks become free of interaction. Quarks inside the individual nucleons forms a colourless matter known as quark matter. \cite{glend} Heiselberg, Pethick and Staubo\cite{heiselberg} proved, in their paper, that the lowest energy state of matter with high density is constituted of quark matter in coexistence with nuclear matter. From this consideration they predicted that in the interior of neutron star finite fraction of constituent matter might be converted into quark matter. It is generally believed that conversion of baryonic matter into quark matter inside the neutron star would result into transient events like neutrino or gamma ray burst \cite{drago}.  However, it is always difficult to discriminate between a hybrid star and a neutron star through observation. Recently, Alford  et al. \cite{alford1} have conformed that hybrid stars can pretend like neutron stars because they have very similar mass-radius rlationship. Recent observation of the binary millisecond pulsar PSR J1614 - 2230 \cite{demor} predicted the mass of the neutron star to be $1.97 \pm 0.04 M_{\odot}$ which was the highest ever mass measured for a neutron star. Another significant paper by Antoniadis et al \cite{antoniadis} have reported after careful calculations that the mass of the pulsar $J0348 + 0432$ is $2.01 \pm 0.04 M_\odot$. These observations rule out the possibility of existence of hyperons or bosons inside the neutron star because the highest mass of the neutron star predicted by the models with the consideration of hyperons or bosons are much smaller than the observed value. Various studies \cite{drago, alford1, alford,alford3, chamel} have reported that  stiffer equation of state is to be considered for the nucleonic matter,  as well as quark matter to achieve the obserbed maximum mass limit of the neutron stars which are believed to be hybrid stars due to their mass. There are numerous proposed models for hybrid star \cite{drago, chamel, hell, fraga, yan}. In the present paper we propose a relativistic model of hybrid star with strange quark matter obeying the MIT bag model equation of state along with normal baryonic matter. In cosmological modeling interacting two fluid model has been extensively used \cite{berger, chimento, lemets, harko}. In their paper, F. Rahaman et al. \cite{rahaman} modelled galactic dark matter halo using interacting bryonic matter and quark matter. We have defined an interaction between the two types of matter inside the hybrid star through interaction equations obtained from the conservation equation. We have explored the physical nature of repulsive and attractive interaction between the two types of matter, and the case of no interaction as well. The plan of the paper is as follows: in the second section we describe the proposed model. Section $3$ persents the results of the equations of the previous section for two different equations of state of the baryonic matter. Section $4$ elaborates various aspects of the results and their physical significance. In section $5$ we testify our model with respect to some standard tests propsed by earlier studies and compare our results with the observed mass and radius of the compact stars. In the last section we summarise our findings with some concluding remarks.

\section{The Model}
The static spherically symmetric spacetime (in geometrical units
$G=1=c$ here and onwards) is taken as
\begin{equation}
ds^2=   e^{\nu(r)} dt^2-e^{\lambda(r)}
dr^2-r^2(d\theta^2+\sin^2\theta d\phi^2). \label{metric}
\end{equation}
The Einstein field equations (EFE) for the above metric are
\begin{widetext}
\begin{eqnarray}
e^{-\lambda} \left[\frac{\lambda^\prime}{r} - \frac{1}{r^2}
\right]+\frac{1}{r^2}&=& 8\pi \left( \rho + \rho_q\right) ,\label{EFE1}\\
e^{-\lambda}
\left[\frac{\nu^\prime}{r}+\frac{1}{r^2}\right]-\frac{1}{r^2}&=&
8\pi \left(p + p_q\right) ,\label{EFE2}\\
\frac{1}{2} e^{-\lambda} \left[\frac{1}{2}(\nu^\prime)^2+
\nu^{\prime\prime} -\frac{1}{2}\lambda^\prime\nu^\prime +
\frac{1}{r}({\nu^\prime- \lambda^\prime})\right] &=&8\pi \left(p + p_q\right) . \label{EFE3}
\end{eqnarray}
\end{widetext}
Here $\rho$ and $p$ are the energy density and pressure of normal baryonic matter and $\rho_q$ and  $p_q$ are the energy density and pressure of the quark matter. 
The study of inheritance symmetry is very useful in searching the natural relation between geometry
and matter through the Einstein equations. The well known inheritance symmetry is the
symmetry under conformal killing vectors (CKV). The conformal
killing equation for (\ref{metric}) becomes
\begin{equation}
L_\xi g_{ik} =\xi_{i;k}+ \xi_{k;i} = \psi g_{ik}.
\end{equation}
where $L$ is the Lie derivative
operator and $\psi$ is the conformal factor. Here vector $\xi$ generates the conformal symmetry and then
the metric $g_{ik}$ is conformally mapped onto itself along $\xi$. It may be noted that neither $\xi$ nor $\psi$
need to be static even though one considers a static
metric~\cite{Bohmer2008a,Bohmer2008b}. The interesting point is that due to this and several other properties CKVs
provide a deeper insight into the space-time geometry connected to
astrophysical and cosmological realm. 
The above equations give the following equations as
\begin{eqnarray}
\xi^1 \nu^\prime &=&\psi,\label{K1}\\
\xi^4  &=& C_1  ,\label{K2}\\
\xi^1 & =& \frac{\psi r}{2},\label{K3} \\
\xi^1 \lambda ^\prime + 2 \xi^1 _{,1} &=&\psi.\label{K4}
\end{eqnarray}
Integration of Eqs. (\ref{K1}-\ref{K2}) yield
\begin{eqnarray}
e^\nu  &=&C_2^2 r^2 , \label{CK1}\\
 e^\lambda  &=& \left(\frac {C_3}
{\psi}\right)^2 ,\label{CK2}  \\   \xi^i &=& C_1 \delta_4^i +
\left(\frac{\psi r}{2}\right)\delta_1^i ,\label{CK3}
\end{eqnarray}
where $C_i$, $i=1,2,3$ are constants of integration. It is interesting to note that the solution of $e^\nu$ in equation (\ref{CK1}) resembles the assumption ($\frac{e^\nu \nu'}{2 r} = constant$)taken by Tolman \cite{tolman} in the solution $IV$ of Einstein's equation. Making use of
Eqs. (\ref{CK1}-\ref{CK3}) in (\ref{EFE1} - \ref{EFE3}), we can write
\begin{eqnarray}
\frac{1}{r^2}\left[1 - \frac{\psi^2}{C_3^2}
\right]-\frac{2\psi\psi^\prime}{rC_3^2}&=& 8\pi \left(\rho + \rho_q\right) ,\label{CFE1}\\
\frac{1}{r^2}\left[1 - \frac{3\psi^2}{C_3^2}
\right]&=& - 8\pi \left(p + p_q\right) ,\label{CFE2}\\
\left[\frac{\psi^2}{C_3^2r^2}
\right]+\frac{2\psi\psi^\prime}{rC_3^2} &=&8\pi \left(p + p_q\right).\label{CFE3}
\end{eqnarray}
Thus Eqs. (\ref{CFE1}-\ref{CFE3}) represent the EFE in terms of the conformal
factor $\psi$.  Since we consider hybrid star model admitting
conformal motion , so all the parameters $ e^{\nu} $, $e^\lambda
$, $ \rho $,$\rho_q$, $p$, $p_q$  could be found in terms of conformal
factor $ \psi $ . In other words, unless one knows the exact form
of $ \psi $, one could not say anything.
The hybrid star is constituted of quark matter at the core within an envelope of normal baryonic matter. We assume an interaction between these two states of matter inside the star. As we have assumed pressure isotropy inside the star, the conservation equation becomes
\begin{equation}
\frac{d(p_{eff})}{dr} + \frac{1}{2} \nu'(\rho_{eff} + p_{eff}) = 0,  \label{CON}
\end{equation}
The equations of state of nuclear matter, as well as the quark matter at low temperature and extremely high density that is relevant to the compact stars, are yet to be confirmed even from the theoretical consideration of Quantum Chromodynamics. We have explored two possible equations of state for the normal baryonic matter. In the first case equation of state of the baryonic matter is taken to be
\begin{equation}
p = m\rho, \label{eos1}
\end{equation} 
and in the second case the equation of state is
\begin{equation}
p = m \rho + b \label{eos12}
\end{equation}
where $m$ is the equation of state parameter and $b$ is another parameter taking values in the range $-\infty < b < \infty$. The MIT bag model is very successful in explaining the observations in particle physics. It is simple, but the researchers have been using it for last three decades to describe the quark matter inside a hybrid star. Following MIT bag equation of state, we have chosen the equation of state for the quark matter as
\begin{equation}
p_q = \frac{1}{3}\left(\rho_q - 4B \right), \label{eos2}
\end{equation}
Where $B$ is the Bag constant being expressed in units of $MeV / (fm)^3$. In literature we get the value of B in the range $60 - 80 MeV / (fm)^3$. Fraga et al. \cite{fraga} have taken $B = (150 MeV)^4$ for zero temperature quark matter in $\beta$ equilibrium. Chamel et al. \cite{chamel}  in their study have used the values of effective Bag constant to be $78.6, 65.5$ and $56.7$ $MeV / (fm)^3$ for the nucleonic equations of state BSk19, BSk20 and BSk21. In the present study we take $B = 83$ $MeV / (fm)^3$ which in geometric units becomes $B = 0.0001$ $(km)^{-2}$.

\section{Results}
From equations (\ref{CFE2}) and (\ref{CFE3}) we get the solutions of $\psi$ as
\begin{equation}
\psi =  \pm \sqrt{2C_{3}^2 + 4Cr^2} \label{psi}
\end{equation}
where $C$ is the constant of integration. However, since a negative value of the conformal factor is not physical, we accept the positive solution of the conformal factor $\psi$.
\begin{equation}
\psi' = \frac{4Cr}{\psi}
\end{equation}
\begin{equation}
\psi \psi' = 4 Cr
\end{equation}
For $\psi = 0$, we get the condition for isometry. From (\ref{psi}), with this substitution one can find out that
\begin{equation}
r = \sqrt{- \frac{C_3^2}{2 C}} \label{motion}
\end{equation}
Equation (\ref{motion}) will give physically acceptable, real solution if $C < 0$. Thus for $C < 0$, the equation gives the radius for which space time shows a isometry. 

\subsection{Case 1: EOS $p = m\rho$}
The density of normal baryonic matter is
\begin{widetext}
\begin{equation}
\rho =  \frac{1}{(3m - 1)(8\pi r^2)} + \frac{3C}{(3m - 1)(2\pi C_{3}^2)} + \frac{4B}{(3m - 1)}
\end{equation}
The density of quark matter is
\begin{equation}
\rho_q = -\frac{3(1-m)}{(3m - 1)(16\pi r^2)} - \frac{9(m+1)C}{(3m - 1)(8\pi C_{3}^2)} -\frac{4B}{(3m - 1)} \label{roq1}
\end{equation}
\end{widetext}
From equation (\ref{CON}) we get the equations for interaction as follows:
\begin{equation}
\rho' + \nu' \frac{(1 + m)\rho}{2 m} = Q, \label{int11}
\end{equation}
\begin{equation}
\rho_q' + 2 \nu'  \left(\rho_q - B \right) = -3 Q, \label{int12}
\end{equation}
Now, substituting the values of $\rho$ and $\rho'$ in (\ref{int11})
we get
\begin{equation}
Q=\frac{1-m}{8\pi r^{3}(3m-1)m}+\frac{\left(m+1\right)}{m(3m-1)r}\left[\frac{3C}{2\pi C^{2}_{3}}+4B\right]
\end{equation}
and substituting the values of $\rho_{q}$ and $\rho^{'}_{q}$ in(\ref{int12}) we get
\begin{equation}
Q=\frac{1-m}{8\pi r^{3}(3m-1)}+\frac{\left(m+1\right)}{(3m-1)r}\left[\frac{3C}{2\pi C^{2}_{3}}+4B\right]
\end{equation}
 Since $Q$ is same in both expression then comparing the coefficient of $\frac{1}{r^{3}}$ and $\frac{1}{r}$,  we get $m = 1$.\\
 
Now for $m=1$, we get
\begin{equation}
Q=\frac{1}{r}\left[\frac{3C}{2\pi C^{2}_{3}}+4B\right]
\end{equation}
In this case three possibilities would occur:\\

\[ 1. ~~ Q > 0 ~~ if ~~ B > -\frac{3C}{8\pi C^{2}_{3}}\] 
\[ 2. ~~ Q < 0 ~~ if ~~ B < -\frac{3C}{8\pi C^{2}_{3}}\] 
\[ 3. ~~ Q = 0 ~~ if ~~ B = -\frac{3C}{8\pi C^{2}_{3}}\] 

Interacting force between normal baryonic matter and strange quark matter is repulsive in nature if $B > -\frac{3C}{8\pi C^{2}_{3}}$ and attractive if $B < -\frac{3C}{8\pi C^{2}_{3}}$.  For $B = -\frac{3C}{8\pi C^{2}_{3}}$, 
there is no interaction between normal baryonic matter and quark matter.\\
  
In the given equation the condition $B > 0$ and $C_3^2 > 0$, we get $-\infty < C < 0$. 

 % For $C_3^2 = 1$ and $B = 0.001$ we get $C = -0.00837758041$ which is in conformity with the afore mentioned condition.

\subsection{Case 2: EOS $p = m\rho + b$}

The density of normal matter is
\begin{widetext}
\begin{equation}
\rho = \frac{1}{8\pi r^2 (3m - 1)} + \frac{3C}{(3m-1)2\pi C_3^2}+\frac{4B}{(3m-1)}-\frac{3b}{(3m-1)} \label{ro2}
\end{equation}
The density of the quark matter is
\begin{equation}
\rho_q =\frac{3(m-1)}{16\pi r^2 (3m - 1)} - \frac{9C(m+1)}{(3m-1)8\pi C_3^2}-\frac{4B}{(3m-1)}+\frac{3b}{(3m-1)} \label{roq2}
\end{equation}
\end{widetext}
The interaction equations for the present case are
\begin{equation}
\rho' + \frac{1}{r}\left(\frac{(m + 1)\rho}{m} + \frac{b}{m}\right) = Q \label{int21}
\end{equation}
\begin{equation}
\rho_q' + 2 \nu'\left(\rho_q - B\right) = - 3Q \label{int22}
\end{equation}
Substituting the values of $\rho$ and $\rho^{'}$ in the interaction equations (\ref{int21})we get
\begin{widetext}
\begin{equation}
Q = \frac{1-m}{8\pi r^{3}m(3m-1)}+\frac{\left[ \frac{3C(m+1)}{2\pi C^{2}_{3}}+4B(m+1)-4b\right]}{m(3m-1)r}
\end{equation}
Substituting the values of $\rho_{q}$ and $\rho^{'}_{q}$ in the interaction equations (\ref{int22}), we get
\begin{equation}
 Q = \frac{1-m}{8\pi r^{3}(3m-1)}+\frac{\left[ \frac{3C(m+1)}{2\pi C^{2}_{3}}+4B(m+1)-4b\right]}{(3m-1)r} 
\end{equation}
\end{widetext}
 Since $Q$ is same in both expressions, comparing coefficient of  $\frac{1}{r^{3}}$ and $\frac{1}{r}$,
we get $m=1$.\\ 

Now for $m=1$, we get
 \begin{equation} 
 Q=\frac{\left[\frac{3C}{\pi C^{2}_{3}}+8B-4b\right]}{2r}
 \end{equation}
As in the earlier case, here also  three possibilities would occurr:\\
 
\[ 1.~~Q > 0 ~~if~~ B > \frac{b}{2} - \frac{3 C}{8 \pi C_3^2}\]
\[ 2.~~Q < 0 ~~if~~ B < \frac{b}{2} - \frac{3 C}{8 \pi C_3^2} \]
\[ 3.~~Q = 0 ~~if~~ B = \frac{b}{2} - \frac{3 C}{8 \pi C_3^2}\]

Interacting force between normal baryonic matter and strange quark matter is repulsive in nature if  $B > \frac{b}{2} - \frac{3 C}{8 \pi C_3^2}$ and attractive if $B < \frac{b}{2} - \frac{3 C}{8 \pi C_3^2}$. For $B = \frac{b}{2} - \frac{3 C}{8 \pi C_3^2}$, there is no interaction between normal baryonic matter and quark matter.\\

%If we put $m=1$ then $B=\frac{b}{2}-\frac{3C}{8\pi C^{2}_{3}}$.

\section{Discussion of the results}
~~~~For $m = 1$, the expresion for $\rho_q$ in equation (\ref{roq1}) boils down to
\begin{equation}
\rho_q = - \frac{9 C}{8 \pi C_{3}^2} -2 B
\end{equation}
The first term in this expression is positive as $C < 0$. However, the second term is negative. So, the energy of the quark matter within the star appears to decrease with increasing $B$. It is interesting that similar observation was pointed out by P.K.Panda et al \cite{panda} in their paper.

~~~~Similarly, for case $II$, the equation (\ref{roq2}) reduces to the form
\begin{equation}
\rho_q =  - \frac{18C + 32 B \pi C_3^2 - 24 b \pi C_3^2}{16 \pi C_3^2} = -\frac{9C}{8 \pi C_3^2} - 2 B + \frac{3 b}{2}
\end{equation}
The role of the Bag constant is similar to the previous case. However, the term $\frac{3 b}{2}$ provides the opposite effect.

~~~~The radius of the star can be found out from the condition $p_{effective}(R) = 0$ where $R$ is the radius of the star. The effective pressure for the first case is 
\begin{equation}
p_{effective} = \rho + \frac{\rho_q}{3} - \frac{4 B}{3} \label{pe1}
\end{equation}
For the afore mentioned condition the equation (\ref{pe1}) gives 
\begin{equation}
R = \sqrt{- \frac{C_3^2}{6 C}} \label{radius}
\end{equation}
For the second case, the $p_{effective}$ is given by
\begin{equation}
p_{effective} = \rho + b + \frac{\rho_q}{3} - \frac{4 B}{3}  \label{pe2}
\end{equation}
The condition $p_{effective}(R) = 0$ gives the expression of radius same as the equation (\ref{radius}). It is interesting to note that the solution for the radius sets a constraint for the constant $C$. Since $C_3^2$ can not be negative, from (\ref{radius}) we get $C < 0$. This is in consistence with the equation (\ref{motion}).

~~~~The mass function of the star is given by
\begin{equation}
M_{eff}(r) = \int_{0}^r 4 \pi r'^2 \left(\rho + \rho_q \right) dr' \\
= \frac{r}{4} - \frac{C r^3}{2 C_3^2} \label{M}
\end{equation}
For $r = R$ (\ref{M}) reduces to
\begin{equation}
M_{eff} = \frac{R}{4} - \frac{C R^3}{2 C_3^2} \label{Ms}
\end{equation}
Substituting $R$ from (\ref{radius}) one can easily get
\begin{equation}
M_{eff} = \frac{1}{3} \sqrt{- \frac{C_3^2}{6 C}} \label{Ms2}
\end{equation}
From (\ref{radius}) and (\ref{Ms2}) it can be verified that
\begin{equation}
\frac{M_{eff}}{R} = \frac{1}{3} \label{buch}
\end{equation}
This value of $\frac{M_{eff}}{R}$ nicely satisfies the Buchdahl's condition for mass radius relationship of a star \cite{buch}.

~~~~In the equation (\ref{psi}) if we substitute $r = R$ and the value of $R$ from equation (\ref{radius}) we immediately get
\begin{equation}
\psi^2(R) = 2 C_3^2 + 4 C R^2 = \frac{4 C_3^2}{3}
\end{equation}
The value of $\psi$ from this equation comes out to be positive. Thus one can note that conformal symmetry holds at the boundary of the star as well.

\begin{figure}
    \centering
        \includegraphics[scale=0.3]{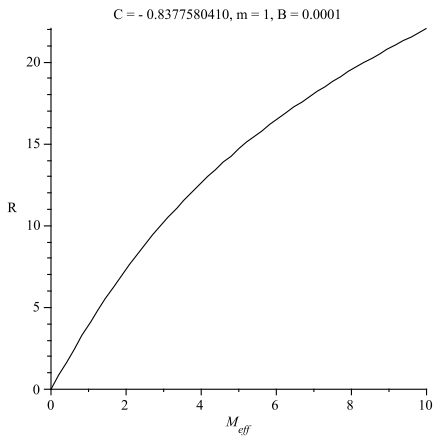}
    \caption{\small{Plot for the variation of $R$  vs $M_{eff}$ (in km) for case $I$.}}
    \label{R_fig}
\end{figure}

If we solve for $R$, in the above equation (\ref{M}) with the substitution of $r = R$, where $R$ be the radius of the star, we get
\begin{widetext}
\begin{equation}
R =\frac{\left(-6 C_3^2 \left( 36 M_{eff} - \sqrt{\frac{6 (216 M_{eff}^2 C - C_3^2)}{C}}\right) C^2 \right)^{\frac{1}{3}}}{6 C} + \frac{C_3^2}{\left(-6C_3^2 \left(36 M_{eff} - \sqrt{\frac{6 (216 M_{eff}^2 C - C_3^2)}{C}} \right) C^2\right)^{\frac{1}{3}}}   \label{R}
\end{equation}
\end{widetext}
The plot for $R$ from equation (\ref{R}) is in fig(\ref{R_fig}).
\begin{figure*}[thbp]
\begin{tabular}{rl}
\includegraphics[width=5.0cm]{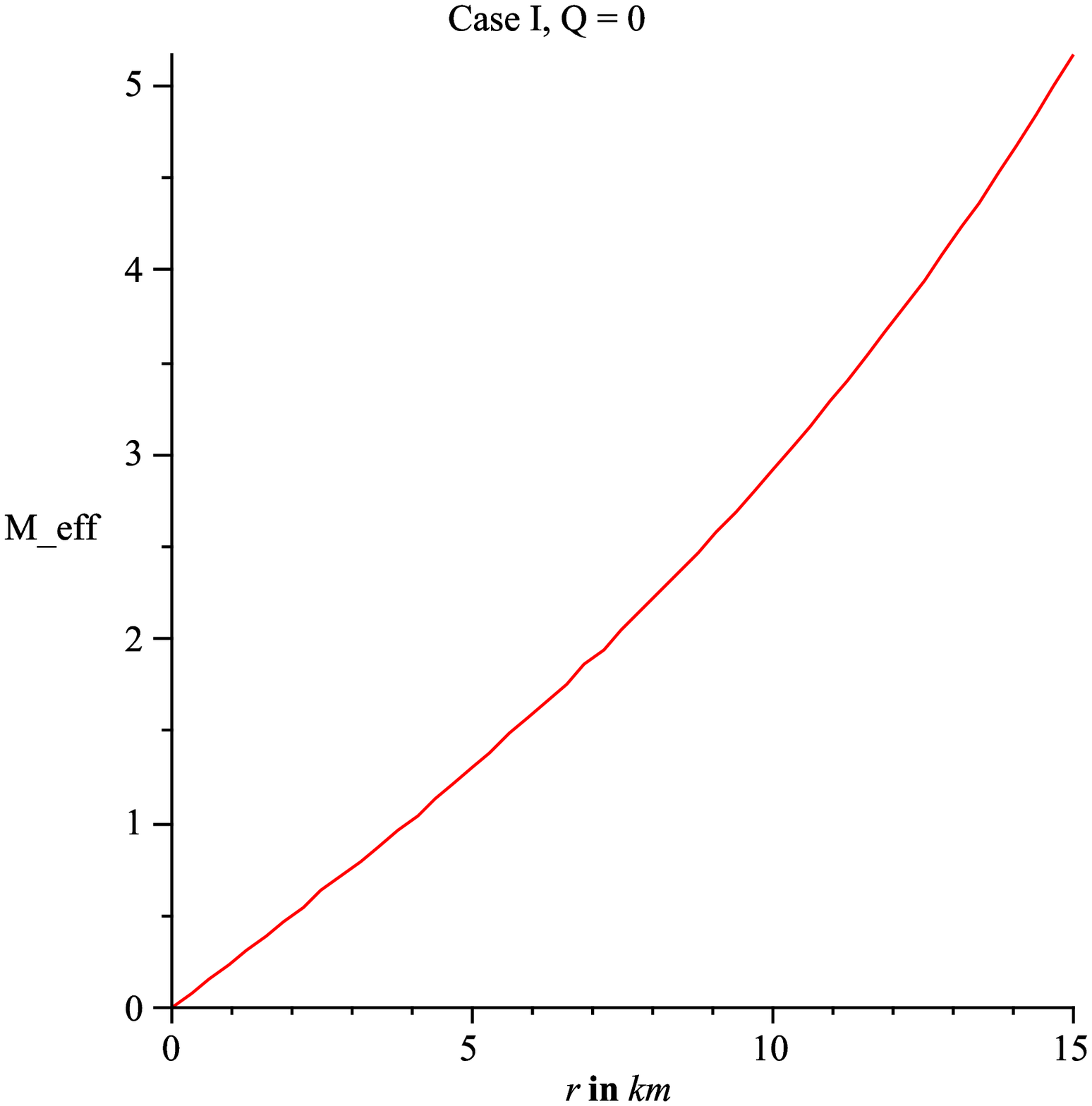}&
\includegraphics[width=5.0cm]{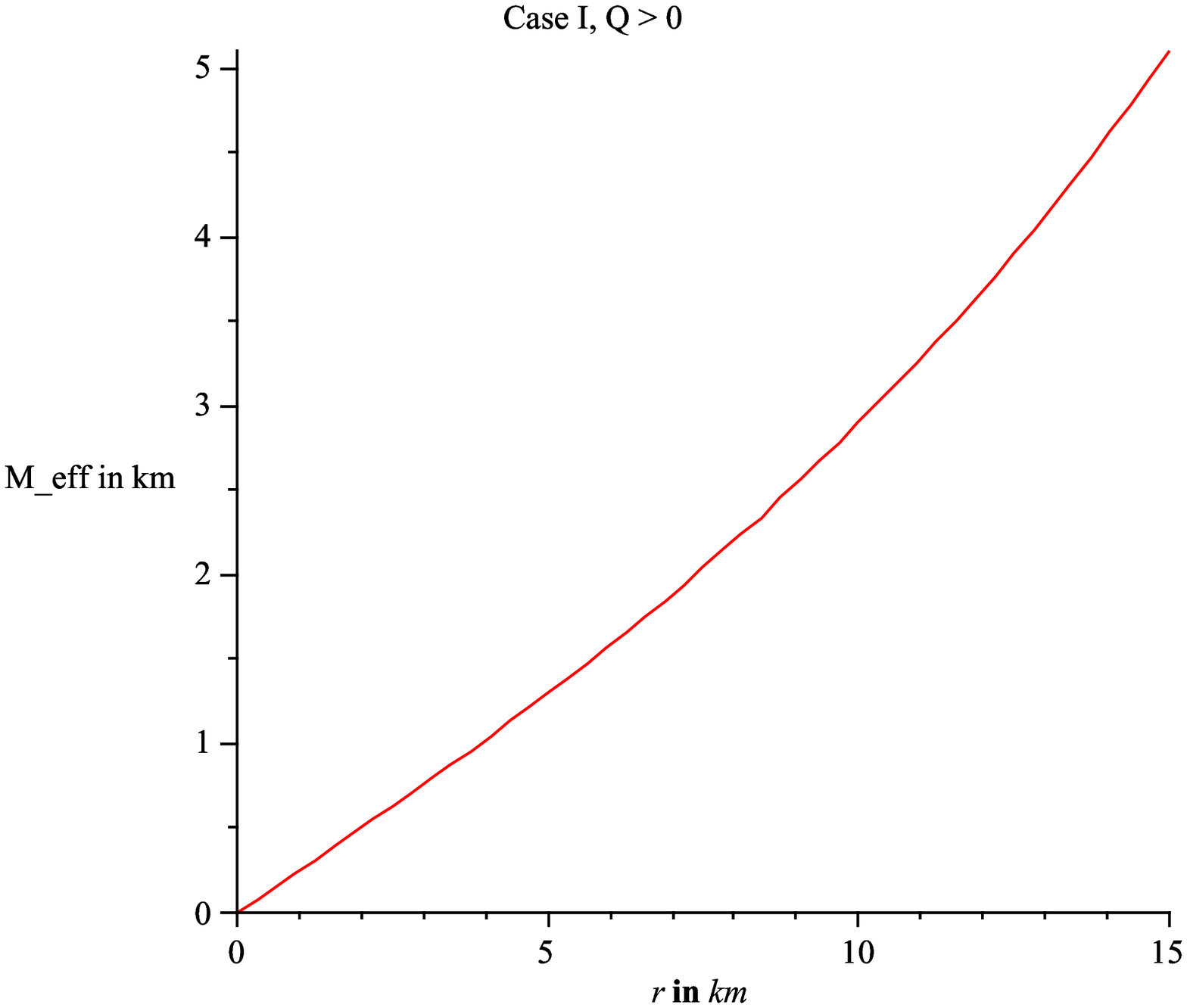}
\includegraphics[width=5.0cm]{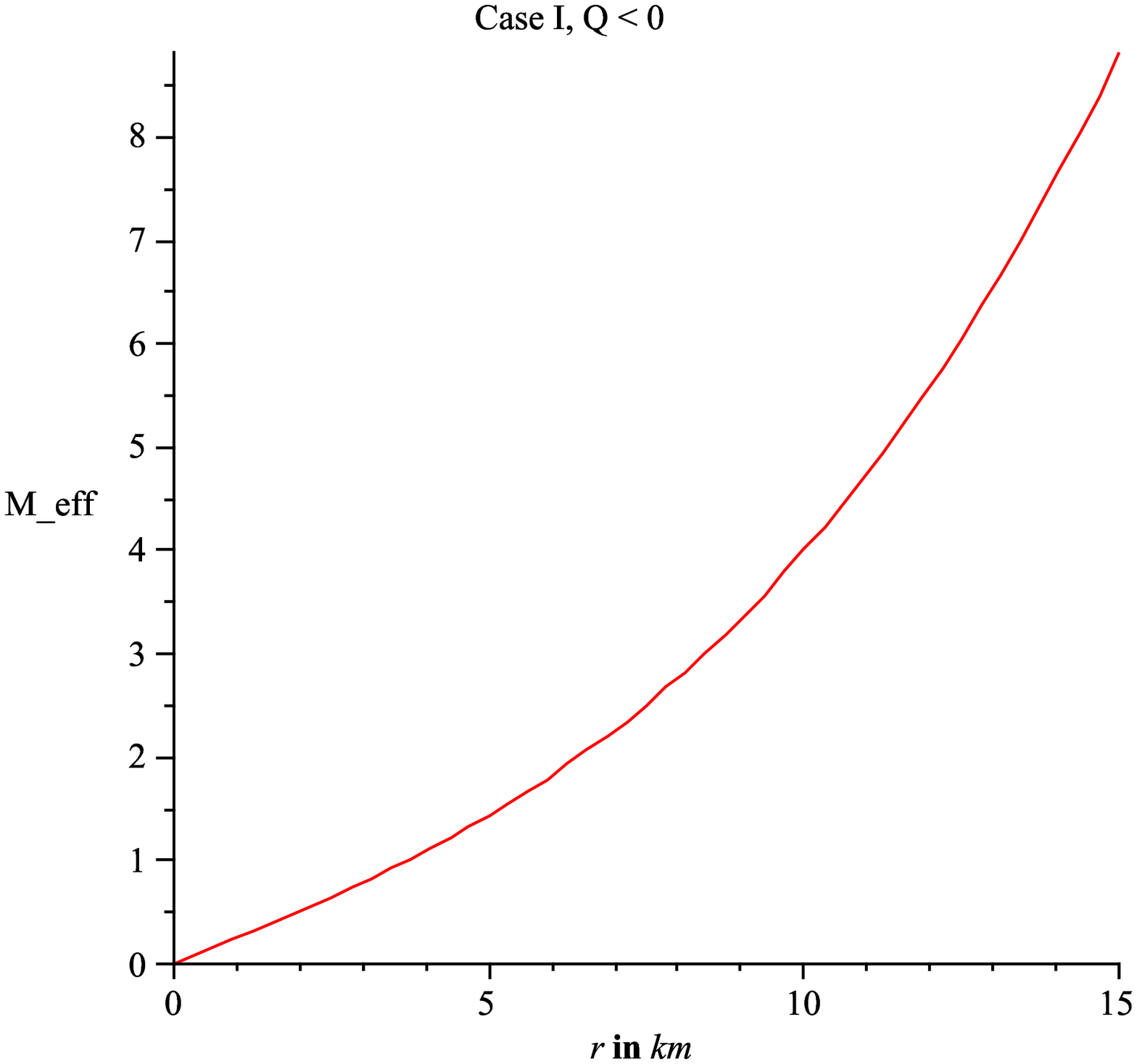}\\
\end{tabular}
\caption{ \small{(Left) Plot for the variation of $M_{eff}(r)$  vs $r$ (in km) for case $I$ $Q = 0$.
        (Middle) Plot for the variation of $M_{eff}(r)$  vs $r$ (in km) for case $I$ $Q > 0$(Right) Plot for the variation of $M_{eff}(r)$  vs $r$ (in km) for case $I$ $Q < 0$.} }
\end{figure*}

\begin{figure*}[thbp]
\begin{tabular}{rl}
\includegraphics[width=5.0cm]{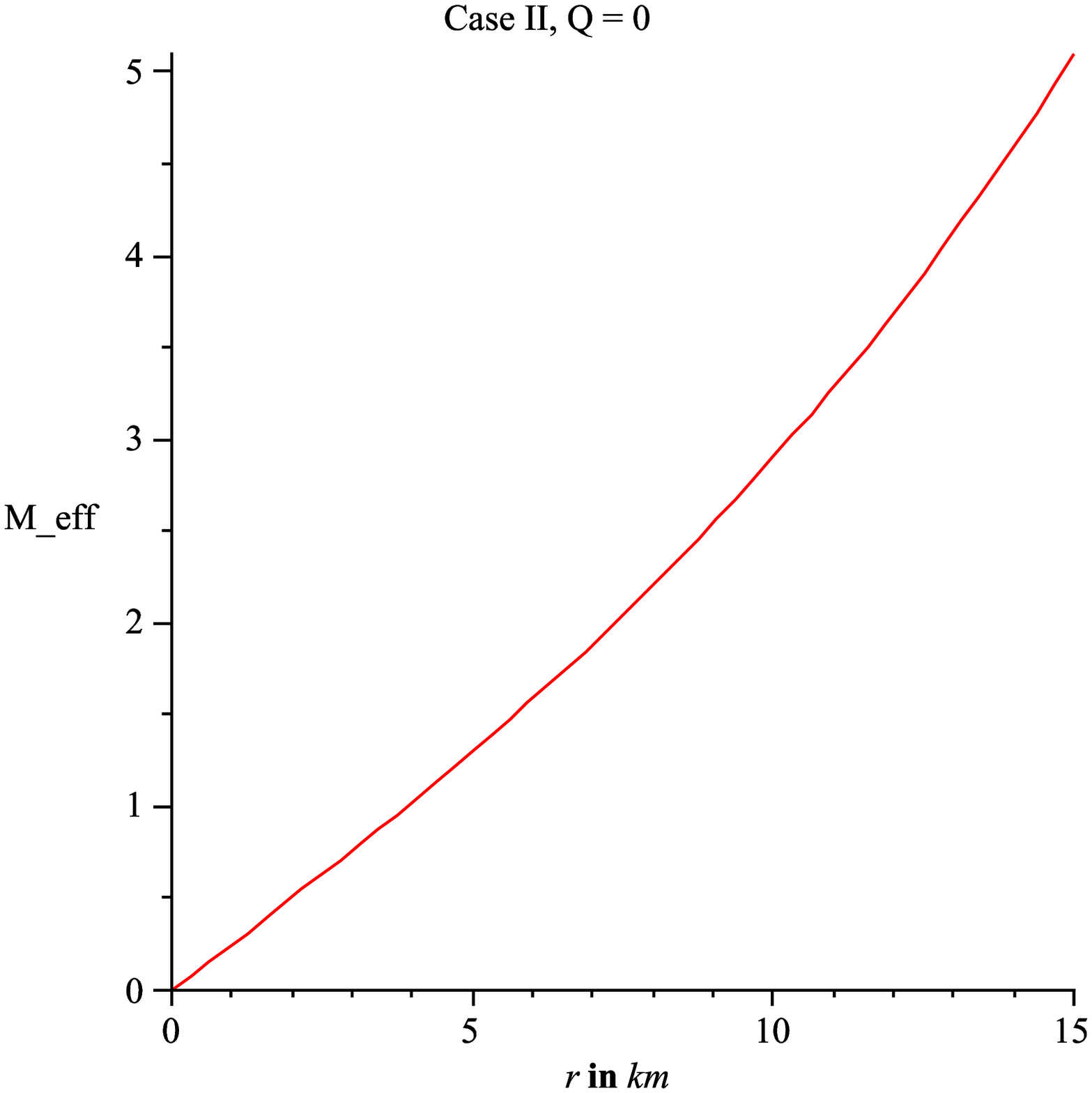}&
\includegraphics[width=5.0cm]{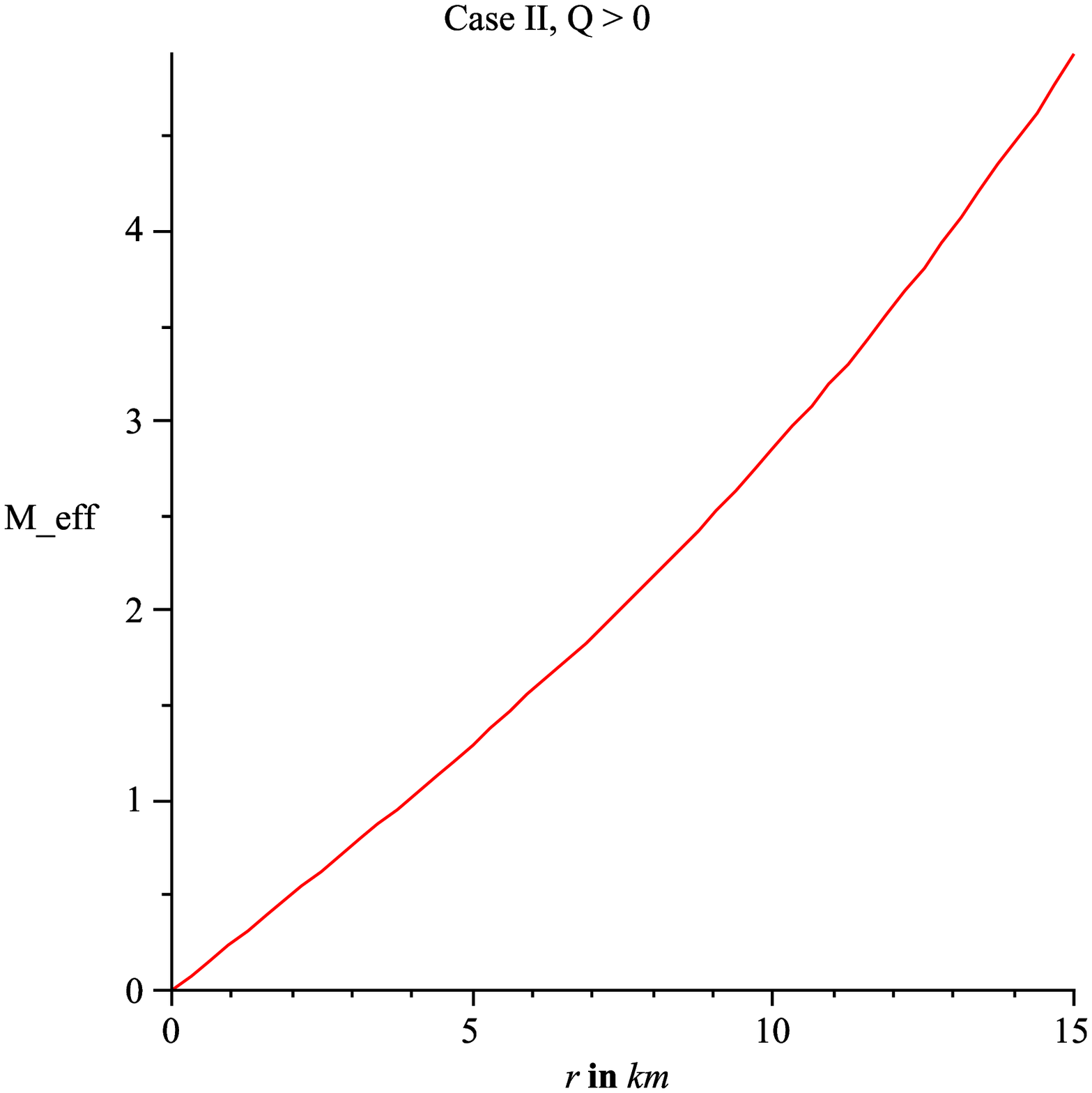}
\includegraphics[width=5.0cm]{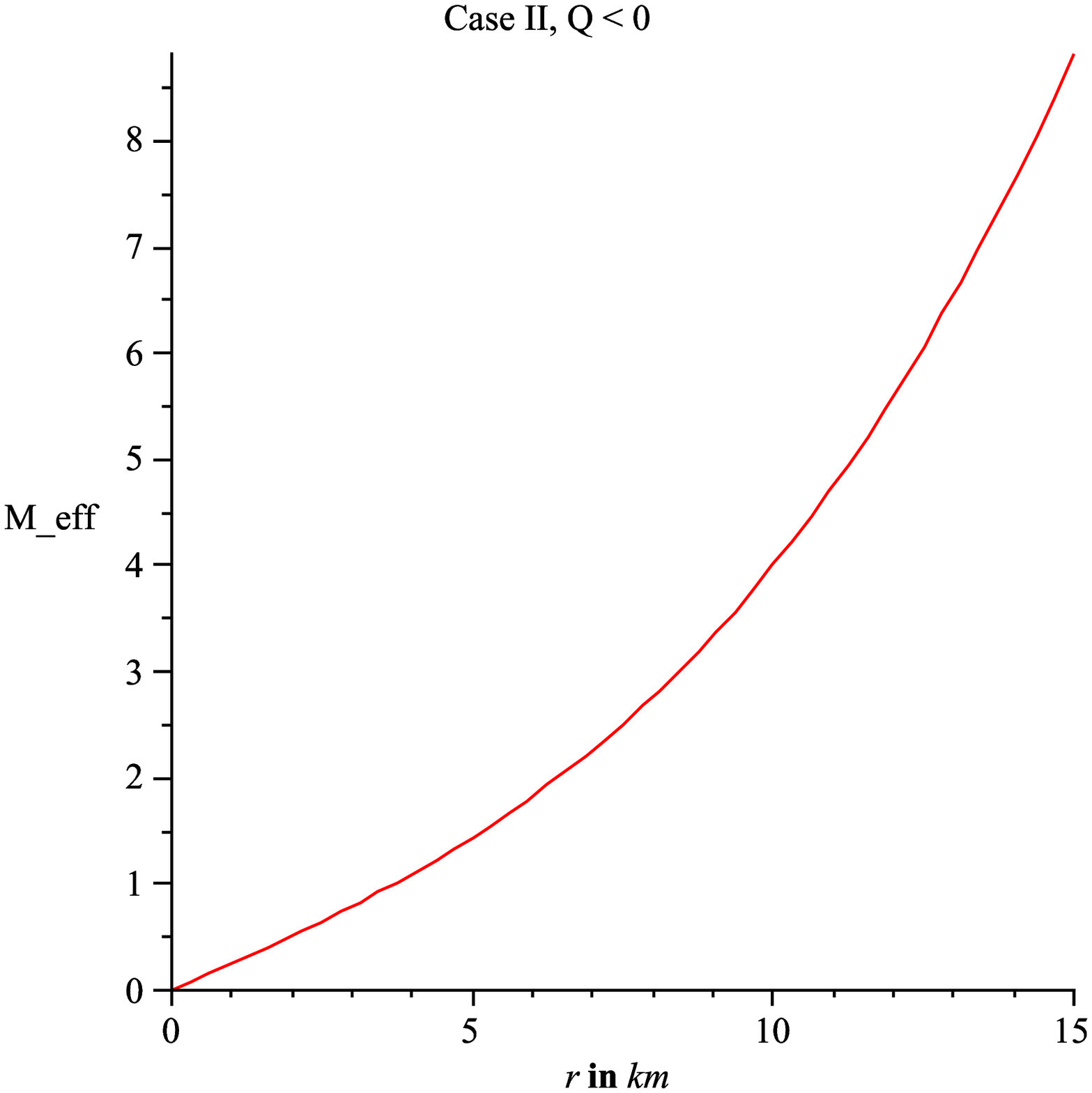}\\
\end{tabular}
\caption{\small{ (Left) Plot for the variation of $M_{eff}(r)$  vs $r$ (in km) for case $II$, $Q = 0$.
        (Middle) Plot for the variation of $M_{eff}(r)$  vs $r$ (in km) for case $II$, $Q > 0$(Right) Plot for the variation of $M_{eff}(r)$  vs $r$ (in km) for case $II$, $Q < 0$.} }
\end{figure*}

 ~~~~The relations for the expressions of Bag constant are particularly interesting because these relations reveal a connection between the constant of particle physics and the constants obtained from the consideration of conformal motion of the hybrid star. Since the constants $C$ and $C_3^2$ are connected to the expression of the conformal factor $\psi$, the equations set up relationship between a microscopic constant and a macroscopic quantity. $C_3^2$ is the constant of proportionality between the quantities $\psi$ and $e^{\lambda}$. 

\begin{figure*}[thbp]
\begin{tabular}{rl}
\includegraphics[width=8cm]{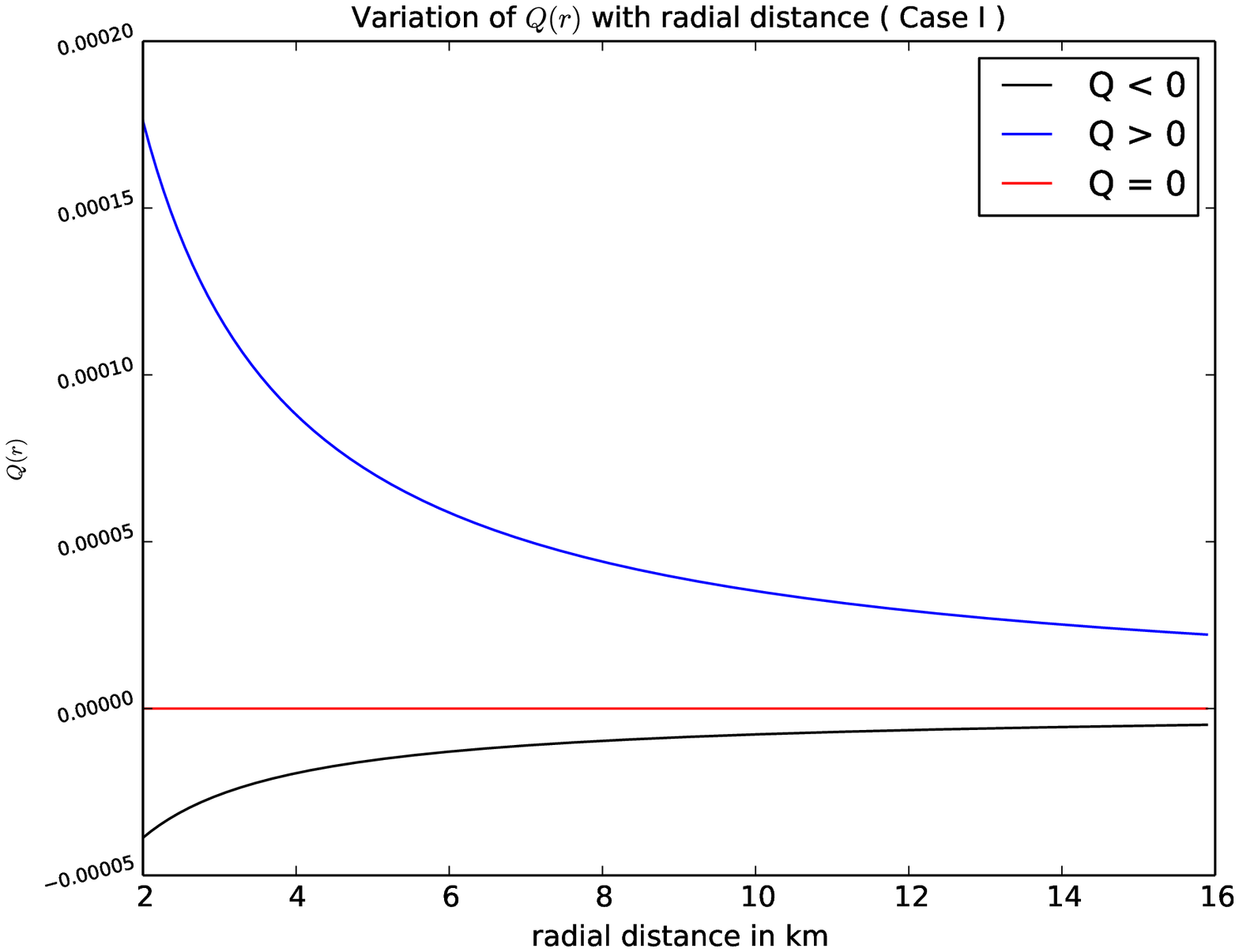}&
\includegraphics[width=8cm]{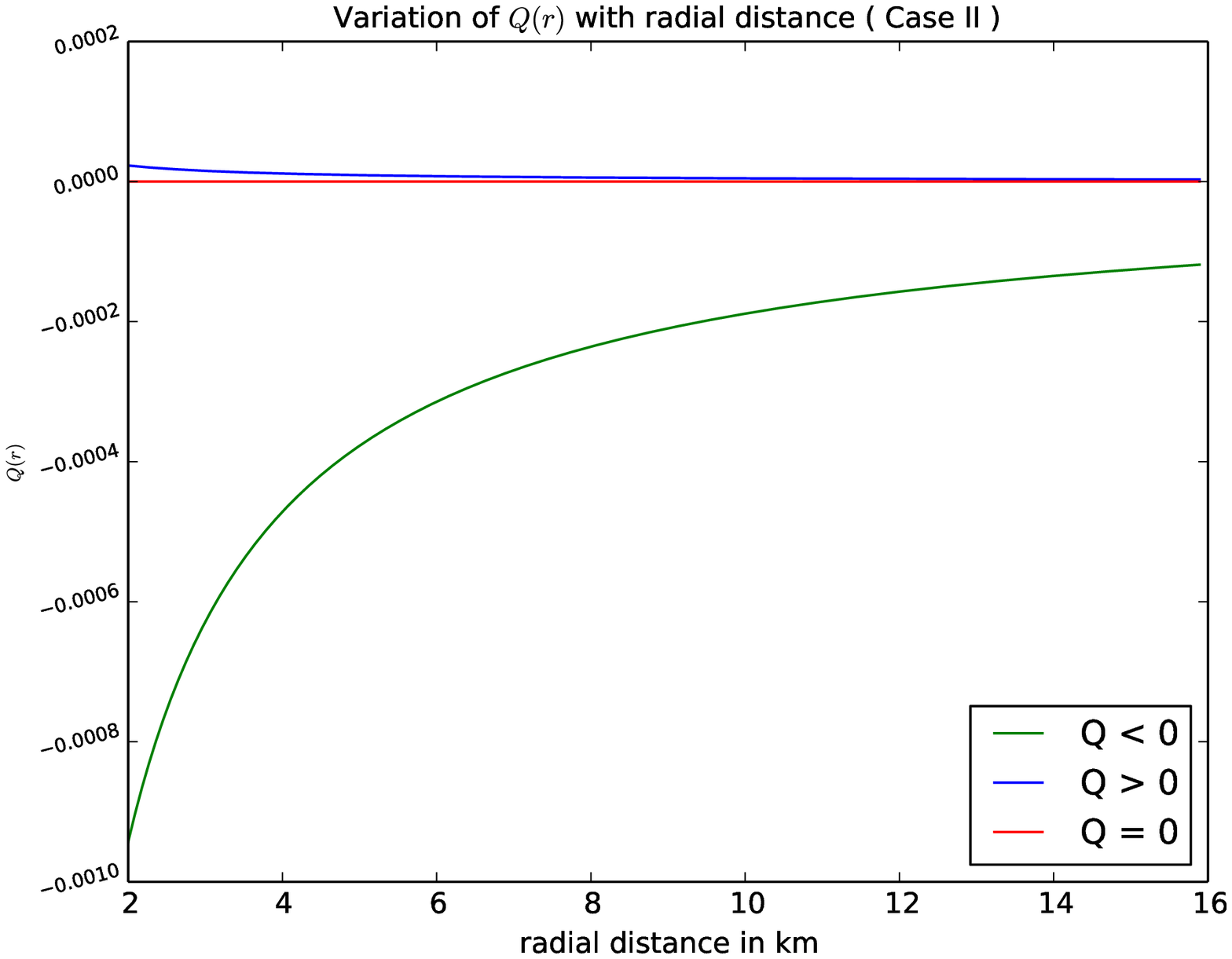} \\
\end{tabular}
\caption{ \small{ (Left) Plot for the variation of $Q$ vs $r$ (in km) for case $I$.
(Right) Plot for the variation of $Q$ vs $r$ (in km) for case $II$.} }
\end{figure*}
\section{Tests of the model}
\subsection{Stability}
~~~~For any realistic equation of state the interior of the star must satisfy the condition $\frac{dp}{d\rho} \leq 1$. This is known as the causality condition. However, the Le Chatelier's principle demands that speed of sound must be real, i.e., $\frac{dp}{d\rho} \geq 0$. Maximally compact stars can be obtained for the consideration $\frac{dp}{d\rho} = 1$ \cite{rhoades}. It is confirmed from various studies \cite{chamel, klahn} that a stiff equation of state for the constituent matter inside the hybrid star has to be considered for explaining the predicted masses of some recently observed compact stars.\cite{demor, antoniadis} Earlier softer equations of state, proposed with the consideration of hyperonization, predicted maximum mass of the star to be much smaller than the recently observed data.\cite{burgio, vidana, schulze}

~~~~For the both cases of our model we find that $\frac{dp_{eff}}{d\rho_{eff}} = 1$. This indicates a very stiff equation of state. However, it has not crossed into the ultrabaric region. For $m = 1$ the equation of state reduces to the form $p = \rho$ which indicates that the energy density would be zero at zero pressure. Since we have taken pressure to be zero outside the star, the density will be zero at the boundary. For the second case, the condition, $m=1$, implies the density will not be zero outside the star. 

~~~~The calculation of Kurkela et al. \cite{kurkela} from the perspective of perturbative Quantum Chromodynamics at zero temperature and also, the non perturbative calculations of QCD at finite temperatures \cite{krasch} show that sound speed inside deconfined quark matter cannot exceed the value $\frac{1}{\sqrt{3}}$. From our model the Bag equation of state gives the speed of sound inside the quark matter to be equal to $\frac{1}{3}$ which is much smaller than the upper limit set by the QCD calculations.

\subsection{Compactness}
The maximum allowable mass-radius relation of the model may be studied on the basis of results obtained by Buchdahl \cite{buch}. Since the trace of the energy momentum tensor for a static spherically symmetric perfect fluid sphere is postulated to be non negative, the quantity $\frac{2 M_{eff}}{R}$ must obey the constraint $\frac{2 M_{eff}}{R} \leq \frac{8}{9}$. The quantity $\frac{2 M_{eff}}{R}$ may be taken as the measure of the compactness of the star. The compactness parameter is plot in figure ($5$). The left figure shows that upto radius of $20$ km the star obeys the Buchdahl condition. However, for case $II$ the condition is maintained upto approximately $6$ km of radius.
\begin{figure*}[thbp]
\begin{tabular}{rl}
\includegraphics[width=5cm]{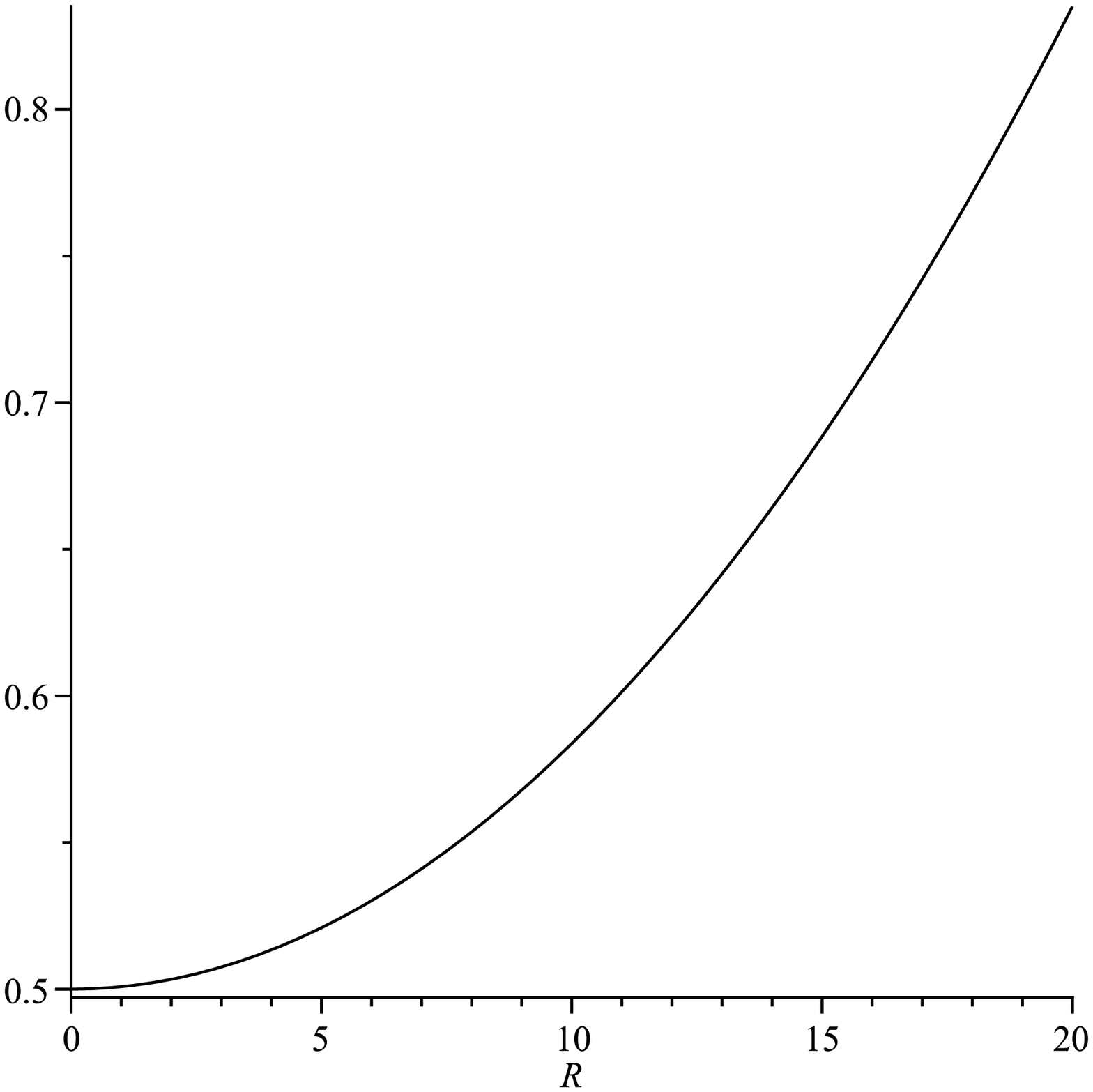}&
\includegraphics[width=5cm]{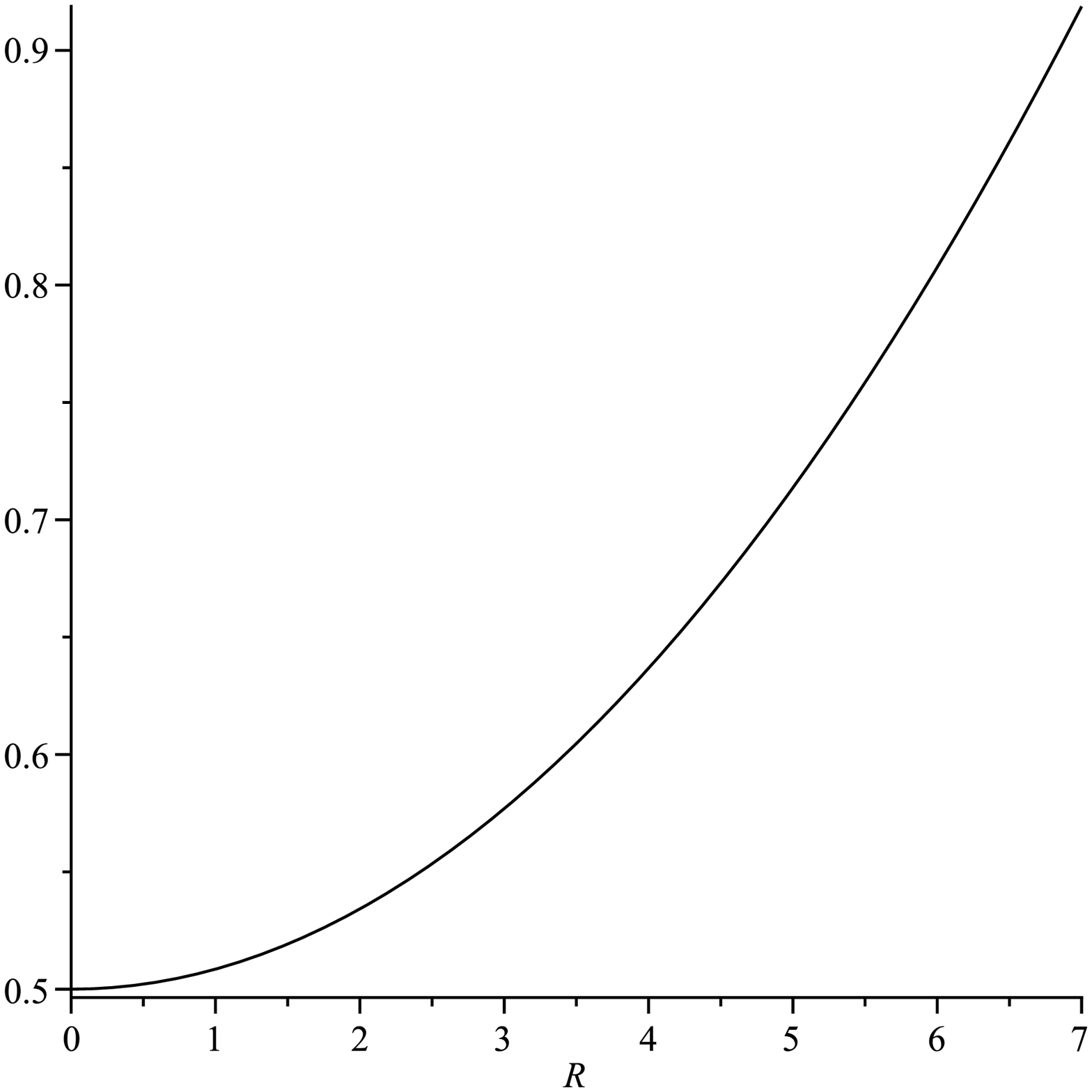} \\
\end{tabular} \label{compactness}
\caption{\small{ (Left) Plot for the variation of $\frac{2 M_{eff}}{R}$ vs $R$ (in km) for case $I$ and $Q = 0$.
(Right) Plot for the variation of $\frac{2 M_{eff}}{R}$ vs $R$ (in km) for case $II$ and $Q = 0$.} }
\end{figure*}

\subsection{Surface gravitational red shift}
The surface redshift of a star is expressed as
\begin{equation}
1 + Z = [ 1 - ( 2 u )]^{- \frac{1}{2}}
\end{equation}
So, the surface redshift is determined by the compactness parameter which we get by substituting the result of the equation (\ref{buch})
\begin{equation}
u = \frac{M_{eff}}{R} = \frac{1}{3} \label{u}
\end{equation}
of the star. The redshift function for the star is given by $Z = 0.732$. From the condition of causality, Lindblom \cite{lindblom} has showed that for nonrotating neutron stars the redshift must be $\leq 0.9$. It can be noted that the upper limit set by the calculations of Lindblom is not crossed for the present model of the hybrid star.

\subsection{Comparison with available data}
\begin{figure}
    \centering
        \includegraphics[scale=0.3]{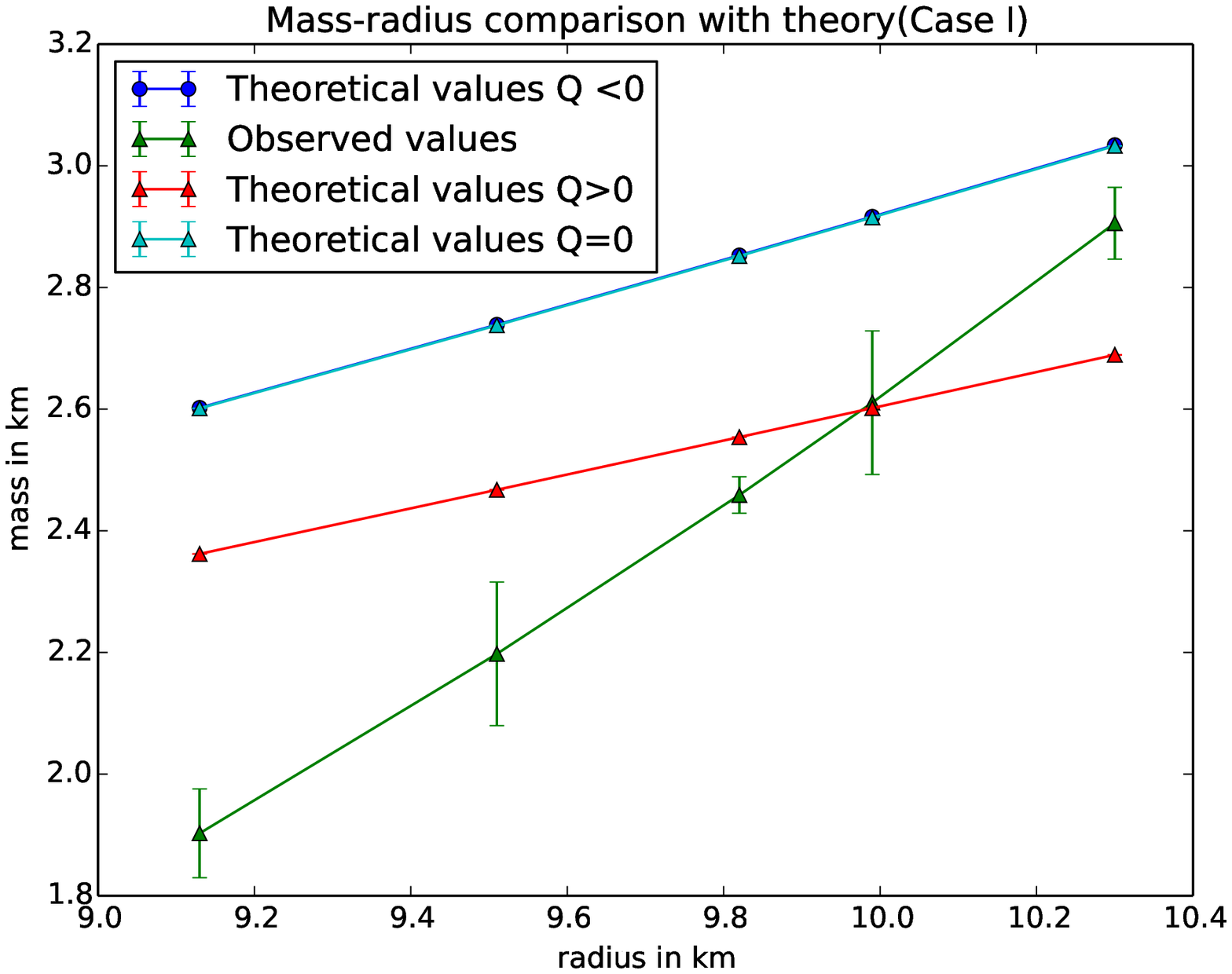}
    \caption{\small{Comparison of calulated values from the model and observed values of $M$ for case$I$.}}
    \label{comparison1}
\end{figure}

\begin{figure}
    \centering
        \includegraphics[scale=0.3]{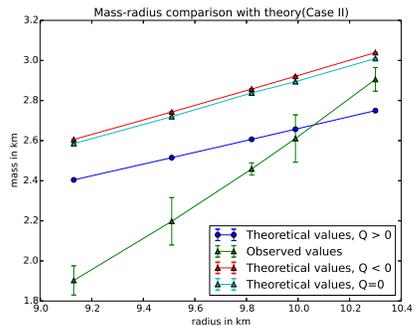}
    \caption{\small{Comparison of calulated values from the model and observed values of $M$ for case$II$.}}
    \label{comparison2}
\end{figure}

Observation from different wavelength ranges of the electromagnetic waves provided fairly accurate data on the masses of the compact stars. However, their radii are not precisely measured. From causality condition constraint, Lattimer and Prakash \cite{lattimer} have given a lower limit to te radii of the compact stars. It is given as
\begin{equation}
R \geq \frac{3.04 M}{R}
\end{equation}
Moreover, from theoretical considerations only Rhoades and Ruffini \cite{rhoades} predicted the maximum mass of the neutron stars to be $3.2 M_{\odot}$. Recent study by Alford et al. \cite{alford} has reported maximum mass of neutron star considering NL3 equation of state for pure neuclear matter to be $2.77 M_{\odot}$ with radius $14.92$ km. In our model, substituting $R = 14.92$ km in the mass function, one may find out that for case I, consideration of repulsive interaction ($Q > 0$) gives the stellar mass $\approx 2.76 M_{\odot}$ which is a fairly good match with the result obtained by Alford et al. For other two cases ($Q < 0 \& Q = 0$) our results violate the upperlimit put by Rhodes and Ruffini ($C_3^2 = 10000 , B = 0.0001$). For case II, repulsive interaction leads to a mass of the star $\sim 2.75 M_{\odot}$ and other two conditions violate the $3.2 M_{\odot}$ limit. Based on Nambu-Jona-Lasinio model, Orsaria et al. \cite{orsaria} have showed that mixed phases of deconfined quark matter and confined hadronic matter possibly exist in neutron star of mass around $2.1 M_{\odot}$ with radii $ \sim 12 - 13$ km. From the data on mass and radius of different compact stars obtained from various studies \cite{demor, freire, rawls}, we have compared the observed data and the data calculated from the present model. The plots are shown in figures (\ref{comparison1}) and (\ref{comparison2}). From the plots, and from the tables (\ref{one}) to (\ref{six}) as well, it can be noted that for different choices of the parameters the plots match closely with the observational data at higher masses and for repulsive interaction between quark matter and normal baryonic matter ($ Q > 0$). The match is appreciable for masses above $2.3$ km. 
\begin{table*}
\centering \caption{The values of the mass and radius for  various
compact stars  [ Case I, $Q < 0$, $B = 0.0001$, $C_3^2 = 10000$, $C = -5$ ]  }

\begin{tabular}{|c|c|c|c|l|} \hline
Compact Stars & Radius ( in km)   & Mass ($M_{\odot}$) &  Mass in km  & Mass from model(km)\\
\hline $PSR J1614-2230$ & $10.3$
 & $1.97 \pm 0.04$ & $2.9057 \pm 0.059$ &~~~$2.7935454$\\
\hline $Vela X - 12$ & $9.99$ & $1.77 \pm 0.08$& $2.6107 \pm 0.118$ &~~~$2.6969006$\\
\hline  $PSR J1903+327$ & $9.82$
 & $1.667 \pm 0.021$& $2.4588 \pm 0.03$ &~~~$2.64439323$\\ \hline $Cen X - 3$  &   $9.51$&$1.49 \pm 0.08$& $2.1977 \pm 0.118$ &~~~$2.54951707$  \\
 \hline $SMC X - 1$ &   $9.13$&$1.29 \pm 0.05$&$1.9027 \pm 0.073$ &~~~$2.4347097$ \\
  \hline\end{tabular} \label{one}
\end{table*}

\begin{table*}
\centering \caption{The values of the mass and radius for  various
compact stars  [ Case I, $Q > 0$, $B = 0.0001$, $C_3^2 = 10000$, $C = -1$ ]  }

\begin{tabular}{|c|c|c|c|l|} \hline
Compact Stars & Radius ( in km)   & Mass ($M_{\odot}$) &  Mass in km  & Mass from model(km)\\
\hline PSR J1614-2230 & $10.3$
 & $1.97 \pm 0.04$ & $2.9057 \pm 0.059$ &~~~$2.62963635$\\
\hline Vela X - 12 & $9.99$ & $1.77 \pm 0.08$&  $2.6107 \pm 0.118$ &~~~$2.54735015$\\
\hline  PSR J1903+327 & $9.82$
 & $1.667 \pm 0.021$ &$2.4588 \pm 0.03$ &~~~$2.50234831$\\ \hline Cen X - 3  & $9.51$ &$1.49 \pm 0.08$& $2.1977 \pm 0.118$&~~~$2.42050427$ \\
 \hline SMC X - 1 &   $9.13$&$1.29 \pm 0.05$& $1.9027 \pm 0.073$ &~~~$2.32055242$ \\
  \hline\end{tabular} \label{two}
\end{table*}

\begin{table*}
\centering \caption{The values of the mass and radius for  various
compact stars  [ Case I, $Q = 0$, $B = 0.0001$, $C_3^2 = 10000$, $C = -2.094395102$ ]  }

\begin{tabular}{|c|c|c|c|l|} \hline
Compact Stars & Radius ( in km)   & Mass ($M_{\odot}$) &  Mass in km  & Mass from model(km)\\
\hline PSR J1614-2230 & $10.3$
 & $1.97 \pm 0.04$ & $2.9057 \pm 0.059$ &~~~$2.6894301$\\
\hline Vela X - 12 & $9.99$ & $1.77 \pm 0.08$ & $2.6107 \pm 0.118$ &~~~$2.60190591$\\
\hline  PSR J1903+327 & $9.82$
 & $1.667 \pm 0.021$ & $2.4588 \pm 0.03$ &~~~$2.55416607$ \\ \hline Cen X - 3  &$ 9.51$ &$1.49 \pm 0.08$ & $2.1977 \pm 0.118$ &~~~$2.46756793$  \\
 \hline SMC X - 1 & $9.13$ & $1.29 \pm 0.05$ & $1.9027 \pm 0.073$ &~~~$2.36219681$ \\
  \hline\end{tabular} \label{three}
\end{table*}

\begin{table*}
\centering \caption{The values of the mass and radius for  various
compact stars  [ Case II, $Q < 0$, $B = 0.0001$, $b =0.00001$, $C_3^2 = 10000$, $C = -8.5$ ]  }

\begin{tabular}{|c|c|c|c|l|} \hline
Compact Stars & Radius ( in km)   & Mass ($M_{\odot}$) &  Mass in km  & Mass from model(km)\\
\hline PSR J1614-2230 & $10.3$
 & $1.97 \pm 0.04$ & $2.9057 \pm 0.059$ &~~~$3.03940898$\\
\hline Vela X - 12 & $9.99$ & $1.77 \pm 0.08$ & $2.6107 \pm 0.118$ &~~~$2.92122627$\\
\hline  PSR J1903+327 & $9.82$
 & $1.667 \pm 0.021$ & $2.4588 \pm 0.03$ &~~~$2.85746062$\\ \hline Cen X - 3  & $9.51$ & $1.49 \pm 0.08$ & $2.1977 \pm 0.118$ &~~~$2.74303627$  \\
 \hline SMC X - 1 & $9.13$ & $1.29 \pm 0.05$ & $1.9027 \pm 0.073$ &~~~$2.60594561$ \\
  \hline\end{tabular} \label{four}
\end{table*}

%\begin{table*}
%\centering \caption{The values of the mass and radius for  various
%compact stars  [ Case II, $Q > 0$ ]  }

%\begin{tabular}{|c|c|c|c|l|} \hline
%Compact Stars & Radius ( in km)   & Mass ($M_{\odot}$) &  Mass in km  & Mass from model(km)\\
%\hline PSR J1614-2230 & 10.3
% & 1.97 \pm 0.04 & 2.9057 \pm 0.059 &~~~2.84818175\\
%\hline Vela X - 12 & 9.99 & 1.77 \pm 0.08& 2.6107 \pm 0.118 &~~~2.74675075\\
%\hline  PSR J1903+327 & 9.82
% & 1.667 \pm 0.021& 2.4588 \pm 0.03 &~~~2.69174154\\ \hline Cen X - 3  &   9.51&1.49 \pm 0.08& 2.1977 \pm 0.118 &~~~2.59252134  \\
% \hline SMC X - 1 &   9.13&1.29 \pm 0.05&1.9027 \pm 0.073 &~~~2.47276212 \\
%  \hline\end{tabular}
%\end{table*}

\begin{table*}
\centering \caption{The values of the mass and radius for  various
compact stars  [ Case II, $Q > 0$, $B = 0.0001$, $b = 0.001$, $C_3^2 = 10000$, $C = -3.2$ ]  }

\begin{tabular}{|c|c|c|c|l|} \hline
Compact Stars & Radius ( in km)   & Mass ($M_{\odot}$) &  Mass in km  & Mass from model(km)\\
\hline PSR J1614-2230 & $10.3$
 & $1.97 \pm 0.04$ & $2.9057 \pm 0.059$ &~~~$2.74983632$\\
\hline Vela X - 12 & $9.99$ & $1.77 \pm 0.08$ & $2.6107 \pm 0.118$ &~~~$2.65702048$\\
\hline  PSR J1903+327 & $9.82$
 & $1.667 \pm 0.021$ & $2.4588 \pm 0.03$ &~~~$2.60651458$\\ \hline Cen X - 3  & $9.51$ & $1.49 \pm 0.08$ & $2.1977 \pm 0.118$ &~~~$2.51511366$  \\
 \hline SMC X - 1 & $9.13$ & $1.29 \pm 0.05$ & $1.9027 \pm 0.073$ &~~~$2.40426776$ \\
  \hline\end{tabular} \label{five}
\end{table*}

\begin{table*}
\centering \caption{The values of the mass and radius for  various
compact stars  [ Case II, $Q = 0$, $B = 0.0001$, $b = 0.00001$, $C_3^2 = 10000$, $C = -7.958701389$]}

\begin{tabular}{|c|c|c|c|l|} \hline
Compact Stars & Radius ( in km)   & Mass ($M_{\odot}$) &  Mass in km  & Mass from model(km)\\
\hline PSR J1614-2230 & $10.3$
 & $1.97 \pm 0.04$ & $2.9057 \pm 0.059$ &~~~$2.9028181$\\
\hline Vela X - 12 & $9.99$ & $1.77 \pm 0.08$ & $2.6107 \pm 0.118$ &~~~$2.7966009$\\
\hline  PSR J1903+327 & $9.82$
 & $1.667 \pm 0.021$ & $2.4588 \pm 0.03$ &~~~$2.73908985$\\ \hline Cen X - 3  & $9.51$ & $1.49 \pm 0.08$ & $2.1977 \pm 0.118$ &~~~$2.63552561$  \\
 \hline SMC X - 1 & $9.13$ & $1.29 \pm 0.05$ & $1.9027 \pm 0.073$ &~~~$2.51081455$ \\
  \hline\end{tabular} \label{six}
\end{table*}

\section{Concluding remarks}
~~~~So, in the present paper we investigate some intricate theoretical questions regarding the physical nature of the hybrid star. The first question is whether the space time of a hybrid star admits conformal symmetry. From our analysis it is evident that the answer is yes. Since conformal motion points to some intrisic symmetry of the space time, we think that the present study would be interesting for further exploration of the nature of the symmetry. Another important problem that we have tried to explore is the nature of interaction between the quark matter and the nucleonic matter inside the hybrid star. However, we did not approach the problem from the point of view of particle physics. From general relativistic considerations and taking both types matter as fluids, we set up interaction equations for the two fluids. The results of the present model fit best with the observational data for $Q > 0$, which implies a repulsive interaction between the quark matter and the nucleonic matter. From the plots (\ref{comparison1}) and (\ref{comparison2}), one can clearly find that the repulsive interaction becomes more and more conspicuous for higher masses. It may be noted that from particle physics also it is predicted that at high densities the mutual repulsions between the nucleonic particles would be dominant. Steinheimer \& Schramm \cite{stein} strongly proposed the existence of repulsive interactions between the deconfined quarks as well as hadrons. Hell \& Weise \cite{hell} in a recent paper pointed out that hybrid matter without the consideration of some strongly repulsive interaction like vector current interaction between quarks, would be very soft. We also get similar finding in the present paper.

~~~~Several models with different predictions are there in the literature to describe cold and dense interacting nuclear matter. Walecka model is one of the most popular models for modeling nuclear matter at high densities. Calculations based on this model show that at high densities for nucleonic matter $P \sim \rho$. Thus, $c_s = \frac{dp}{d\rho} = 1$ \cite{schmitt}. In the present model, we have obtained $m = 1$ for equation of state of the normal baryonic matter. Therefore, from our model also we get $c_s = 1$ which is in concurrence with the prediction of Walecka model.

~~~~A. Chodos et al.\cite{chodos} proposed a strongly interacting particle as a finite region of space where the fields are confined. In their paper they conjectured Bag constant as a finite region having a constant energy per unit volume. Thus this constant must contribute to energy momentum tensor of the star. This in turn infuences the geometry of the space time. In the present model we have got a relation between the Bag constant and the geometry of the space time of the hybrid star through the three relations which we have obtained from the interaction equations for both cases of the equations of state. Since Bag constant is a constant measurable in terrestrial laboratories, we used the measured value of $B$ to find out the possible nature of the space time geometry through the constant $C_3^2$ which is the constant of proportionality between $e^{\lambda}$ and $\psi$, and $C$.

~~~~We have obtained the values of mass and radius of the hybrid star in equations (\ref{radius}) and (\ref{Ms2}) from theoretical calculations. It is intersting to note that the values of the mass and radius satisfies the Buchdahl's condition. This is a very strong support for our model. Moreover, the value of mass calculated from the mass function, obtained from the theory closely match  the mass obtained by Alford et al. \cite{alford} for $Q > 0$.

~~~~However, in the present model we have not taken into account the particle nature of quarks and baryons and considered both as two different fluids inside the hybrid star. The study may be extended considering more realistic equations of state obtained from QCD simulations. \cite{alford}

\section*{Acknowledgments} FR and KC are thankful to the authority of
Inter-University Centre for Astronomy and Astrophysics, Pune,
India for providing them Visiting Associateship under which a part
of this work was carried out. KC is thankful to the University Grants Commision for providing financial support in MRP under which this research work was carried out. AM is thankful to DST for providing financial support under INSPIRE programme.

\end{document}